\documentclass{article}

\usepackage{arxiv}

\usepackage[utf8]{inputenc} 
\usepackage[T1]{fontenc}    
\usepackage{hyperref}       
\usepackage{url}            
\usepackage{booktabs}       
\usepackage{amsfonts}       
\usepackage{nicefrac}       
\usepackage{microtype}      
\usepackage{lipsum}
\usepackage{graphicx,subfigure}
\usepackage{caption}
\title{Efficient Spatial Anti-Aliasing Rendering \\ for Line Joins on Vector Maps}

\author{
  Chaoyang He\thanks{This work was done when Chaoyang He was an staff software engineer at Tencent (www.tencent.com).}  \\
  Department of Computer Science\\
  University of Southern California\\
  Los Angeles, CA \\
  \texttt{chaoyang.he@usc.edu} \\
     \And
  Ming Li \\
  Department of Maps Platform\\
  Tencent\\
  Shenzhen, China\\
  \texttt{littileli@tencent.com} \\
}

\begin{document}
\maketitle

\begin{abstract}
The spatial anti-aliasing technique for line joins (intersections of the road segments) on vector maps is exclusively crucial to visual experience and system performance. Due to limitations of OpenGL API, one common practice to achieve the anti-aliased effect is splicing multiple triangles at varying scale levels to approximate the fan-shaped line joins. However, this approximation inevitably produces some unreality, and the system rendering performance is not optimal. To circumvent these drawbacks, in this paper, we propose a simple but efficient algorithm which uses only two triangles to substitute the multiple triangles approximation and then renders a realistic fan-shaped curve with alpha operation based on geometrical relation computing. Our experiment shows it has advantages of a realistic anti-aliasing effect, less memory cost, higher frame rate, and drawing line joins without overlapping rendering. Our proposed spatial anti-aliasing technique has been widely used in Internet Maps such as Tencent Mobile Maps and Tencent Automotive Maps.
\end{abstract}


\section{Introduction}
In the setting of the smartphone operating system (Android, iOS) and the automotive operating system (Linux, Android, QNX), digital vector maps, a significant component determines the visual experience and system performance in geographic information system (GIS), is built based on OpenGL ES library. It uses points, lines and polygons (areas) as vector data to render real-world map objects such as POIs (Point of Interest), roads, parks, rivers and houses. During the OpenGL rendering process, line joins are used to render the road segment intersections, as shown in Figure 1 (a). The anti-aliasing technique for these line joins is exclusively important since the maps is mainly constituted by road networks, and map users rely on them for route navigation. A better anti-aliasing performance can improve the level of visual verisimilitude and simultaneously optimize the system performance, including memory overhead and rendering frame rate. 

Drawing lines in OpenGL is unsatisfactory. First, the $\textit{GL\_LINES}$ API \cite{noauthor_primitive_nodate} has many limitations: it does not support line joins, line caps, non-integer line widths, widths larger than 10px, or varying widths, which makes it unsuitable for high-quality vector maps. Additionally, OpenGL’s anti-aliasing (multisample anti-aliasing \cite{noauthor_multisampling_nodate}) is not reliably present on all devices and generally is of poor quality. As an alternative, for vector maps rendering, one common approach to achieve the anti-aliased effect is splicing multiple triangles at varying scale levels to approximate the fan-shaped line join and then fitting these triangles with texture or color with gradients at the edge. As shown in Figure 1 (b), edge gradients in multiple triangles jointly provide the anti-aliasing effect for line joins. However, one defect of this approach is that the approximation inevitably produces some unreality, which does not achieve perfect line join arc. Moreover, the system rendering performance is not optimal. When there are many road intersections on the screen, rendering multiple triangles requires multiple draw calls, which lowers the frame-rate and requires much memory to cache vertices. There also exists overlapping rendering in the intersection part of two line segments.

\begin{figure}[h]
\centering     
\subfigure[A line join, rendering the intersection of road segments]{\includegraphics[width=0.45\textwidth]{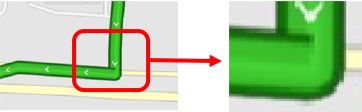}}
\subfigure[The common method to render the line joins]{\includegraphics[width=0.50\textwidth]{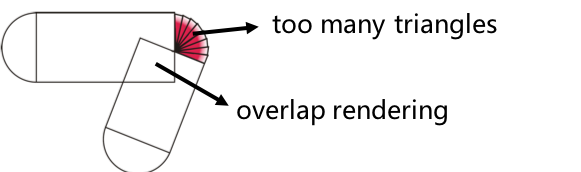}}
\caption{Line Joins on Vector Maps}
\end{figure}
In this paper, we propose a simple but efficient algorithm to address these drawbacks. Our main idea is using only two triangles to substitute the multiple triangles approximation and then render a realistic fan-shaped curve with alpha operation based on geometrical relation computing.  The main advantages of our method are: 1. It implements anti-aliasing with real curve, creating smooth effect for line joins. 2. It significantly decreases the number of triangles to form line joins, which means less draw calls, leading to less memory cost and higher frame rate. 3. The simple design of this method perfectly implement line joins without overlapping rendering.

\section{Method}
\subsection{Preliminaries}
\begin{figure}[h]
  \centering
  \includegraphics[width=0.95\textwidth]{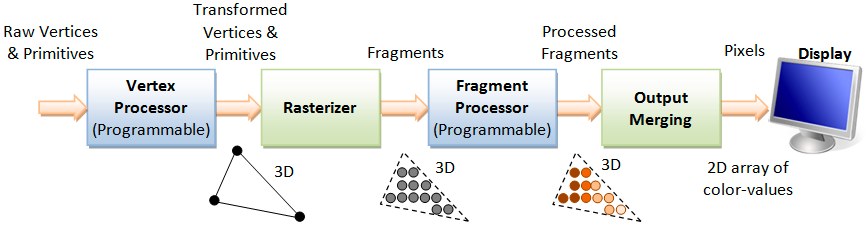}
  \caption{OpenGL Rendering Pipeline}
  \label{fig:pipeline}
\end{figure}

The OpenGL Pipeline is shown in Figure \ref{fig:pipeline}. It consists of the following main stages\cite{noauthor_3d_nodate}: (1) Vertex Shader (Vertex Processing): Process and transform individual vertices. These vertices define the boundaries of primitives which are basic drawing shapes like triangles, lines, and points. (2) Rasterization: Convert each primitive (connected vertices) into a set of fragments. A fragment can be treated as a pixel in 3D spaces, which is aligned with the pixel grid, with attributes such as position, color and texture. (3) Fragment Shader (Fragment Processing): Process individual fragments.
(4) Output Merging: Combine the fragments of all primitives (in 3D space) into 2D color-pixel for the display.

To be more specific, with regard to the process of rendering line segments onto the screen, a line segment is split into triangles, given its width, geographic coordinates of the triangle and anti-aliasing attribute values. Then, vertex data of the triangles including geographic coordinates, anti-aliasing properties are transmitted to vertex shader, creating memory on GPU for storing our vertex data. It follows that each triangle is rasterized into triangle fragments and passed to the fragment shader for processing, in which the anti-aliasing operation is performed. Finally, the GPU combines the results of the previous procedure to render it on the screen.

Vertex Processing and Fragment Processing are programmable. The language in which shaders are programmed depends on the target environment \cite{noauthor_shader_2019}. The official OpenGL and OpenGL ES shading language is OpenGL Shading Language, also known as GLSL. In this work, to optimize spatial anti-aliasing rendering for line joins on vector maps, our algorithm runs on the OpenGL rendering pipeline through shader programming (vertex shader and fragment shader).

\subsection{Algorithm}
\begin{figure}[h]
\centering     
\subfigure[]{\includegraphics[width=0.45\textwidth]{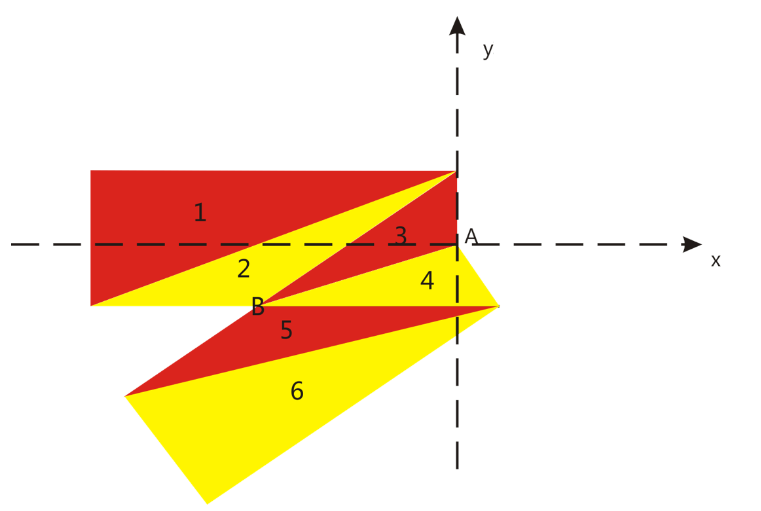}}
\subfigure[]{\includegraphics[width=0.45\textwidth]{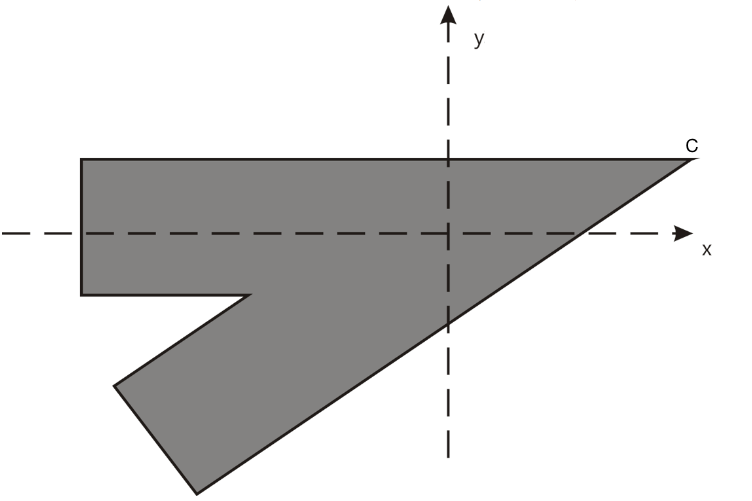}}
\subfigure[]{\includegraphics[width=0.45\textwidth]{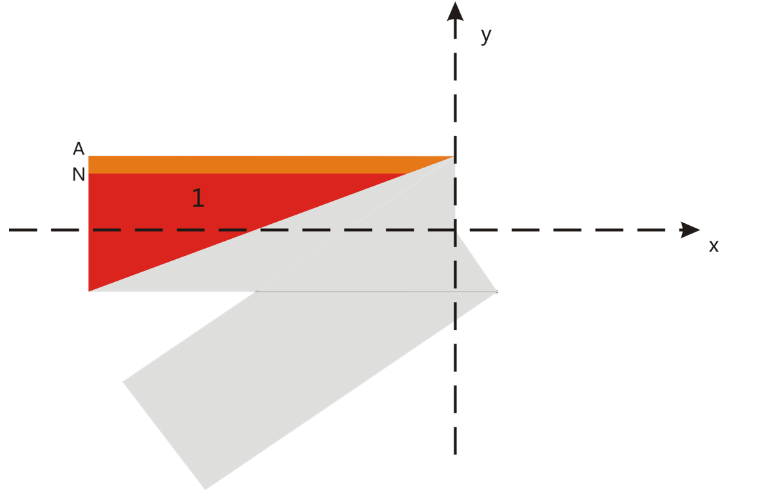}}
\subfigure[]{\includegraphics[width=0.45\textwidth]{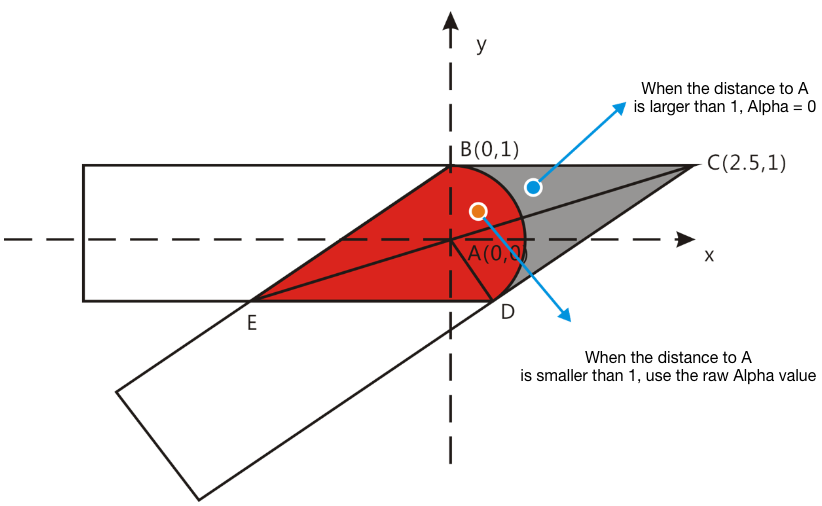}}
\caption{Geometric Schematic of the Proposed Spatial Anti-aliasing Rendering Algorithm}
\label{fig:commonpractice}
\end{figure}

For the main body of the line segment, as shown in Figure \ref{fig:commonpractice} (a), a line segment parallel to the X-axis is denoted as line segment $E$, a unit vector parallel to $E$ and oriented the x-axis direction is denoted as $a$, and another line segment oblique to the X-axis is denoted as line segment $F$, unit vector parallel to $F$ but in the opposite direction to the x-axis is referred as vector $b$. Given the line width of $W$, the length of half a line segment is $\frac{W}{2}$. Here, by adding $(-1)*a*\frac{W}{2}$ to $b*\frac{W}{2}$, the vector $\vec{AB}$ can be easily obtained. Then, in the case where the coordinates of the vertices are known, the two line segments can be split into triangles with number ranging from 1 to 6, and the subsequent triangle fragments can be directly rendered.

For the common practice, the line join is often fitted by multiple triangles, which adds a lot of extra triangle overhead, and the arc is not natural. Different from this method, our proposed method utilizes the geometric relation that the coordinate of the point C in the intersection extension line of $E$ and $F$ can be calculated. Then, Figure \ref{fig:commonpractice} (b) is achieved.


Based on  Figure\ref{fig:commonpractice} (b), for the main part of the line segments, anti-aliasing is implemented by alpha gradient processing. For triangle 1, a threshold N is defined below point A, and alpha gradient processing is performed in the orange region, while the rest retains alpha = 1. Since alpha gradient processing is only performed in Y direction, the pre-stored anti-aliasing property is assigned to the vertex triangle, wherein the X-axis direction attribute is 0, and the Y-axis direction attribute is set to -1, 1 respectively, where 1 represents Y-axis attribute of point A and B. Thus, the triangle segments will only have alpha interpolation in the Y-axis direction.

For the line joins, as shown in figure, the anti-aliasing attribute pre-stored in the vertex triangle, e.g. triangle ABC, is also assigned. Point A, B, C are assigned to (0, 0), (0, 1), (N, 1) respectively, where N dependents on the corner angle. In the rasterization process fragment shader, all fragments which have the distance of 1 to point A form an arc. In fragment shader, we also define an anti-aliasing threshold $N$ which takes value between 0 and 1. The anti-aliasing attribute of the fragments can be assigned by calculating their distance $d$ to point A.
If $d < N$, then its alpha value is 1 which means opaque.
If $N<d<1$, then the fragment is linearly interpolated by d, its alpha value ranges from 0 to 1 which means the gradient transparent; If $d>1$, then its alpha is 0, which means transparent. Finally, for the intersection of line segments, it display as a real arc.

Therefore, in the vector map rendering scene, for the above algorithm steps, spatial anti-aliasing rendering effect for line joins can be achieved through the vertex shader and fragment shader algorithm.

\section{Experiment}
\begin{figure}[h]
  \centering
  \subfigure[Figure A]{\includegraphics[width=0.49\textwidth]{overlap_rendering.png}}
\subfigure[Figure B]{\includegraphics[width=0.49\textwidth]{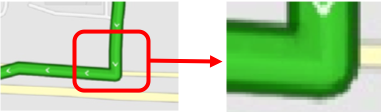}}
  \caption{Rendering Effect of the Proposed Method}
  \label{fig:improved rendering effect}
\end{figure}

We tested the performance of the existing algorithm and our proposed algorithm on the mobile map. In terms of the visual effect, as shown in the figure \ref{fig:improved rendering effect}, we can find that the improved algorithm avoids the problem of repeated corner drawing and the line join is more realistic. 

In terms of system performance, the optimization of the draw call number can reduce the GPU memory and increase frame rate. We selected 10 road networks with different density in Beijing, China for rendering testing, recording their draw calls and comparing the average value. The original method has different numbers of triangles interpolated because of different corner angles, so the number of draw calls in each scene is different. Experiments show that our algorithm has fewer draw calls.
\begin{table}[h]
  \caption{System Performance Comparison}
  \label{table:statistic}
  \centering
  \begin{tabular}{ccc}
    \toprule
 Method & Average Draw Calls \\
\midrule
Fitting Line Joins with Triangles & 33.7 \\
Our Proposed Method & 12 \\ 
    \bottomrule
  \end{tabular}
\end{table}
\section{Conclusion}
we proposed a simple but efficient algorithm to optimize the spatial anti-aliasing rendering. It uses only two triangles to substitute the multiple triangles approximation and then renders a realistic fan-shaped curve with alpha operation based on geometrical relation computing. We conduct experimental evaluation on the real-world mobile map system. It shows our algorithm has advantages of a realistic anti-aliasing effect, decreasing the number of triangles, leading to less memory cost and higher frame rate, and drawing line joins without overlapping rendering.

\bibliographystyle{IEEEtran}
\bibliography{IEEEabrv,ComputerGraphics}  

\end{document}